\documentclass[pre,showpacs]{revtex4}
\usepackage{setspace}
\usepackage{graphicx}
\usepackage{dcolumn}

\begin{document}

\title{Conformational transitions of polyelectrolytes in poor solvents}
\author{Souvik De and Arti Dua}
\affiliation{Department of Chemistry, Indian Institute of Technology, Madras, Chennai-600036, India}

\date{\today}

\begin{abstract}
Starting with a model Hamiltonian, we study using the uniform expansion method conformational behavior of polyelectrolytes in the presence and absence of salt. The uniform expansion method yields all the important local length scales in the polyelectrolyte: the electrostatic blob size at large fraction of charges, the thermal blob size  at low fraction of charges and the sizes of pearls (beads) and necklaces (strings) at intermediate fraction of charges. In the presence of salt, the electrostatic blob size depends on the ionic strength and increases with the increase in the salt concentration. We determine the salt concentration at which the pearl necklace intermediates dissolve and the nature of the transition changes from discontinuous to continuous. This critical salt concentration corresponds to the length scale where the Debye screening length is of the order of the necklace length. 

\end{abstract}


\maketitle

\section{Introduction}
\noindent


Polyelectrolytes are charged polymer with ionizable groups that dissociate ions in solution, leaving the charges of one sign bound to the chain and counterions in solution \cite{t}.  Inspite of the hydrophobic backbone, the presence of the charged side groups make these molecules water soluble, a property which is very important for biological and technological applications \cite{h,b}. The competition between the short range attractive interactions due to poor solvent conditions and the long range repulsion between like charges gives rise to structures not seen in neutral polymers.  The rich conformational behaviour of polyelectrolytes can be finely tuned by changes in the ionic strength, pH,  temperature and condensation agents \cite{expt3}.  However, the presence of the long-range electrostatic interactions in combination with the polymer elasticity make these system difficult to handle both analytically and computationally. 

Simulations on polyelectrolytes in poor solvents show a first order transition from a globular state at low fraction of charges to an extended state at large fraction of charges. The transition proceeds via the formation of metastable pearl-necklace intermediates - locally collapsed globules joined by narrow strings. Pearl-necklace structures have been seen in simulations both in the absence \cite{burkhard, stoll} and presence of counter-ion condensation \cite{k1,k2,k4,k3}. The splitting of a charged globule to form a pearl-necklace structure is similar to that of the Rayleigh instability of a charged liquid droplet, where system gains energy by splitting into smaller droplets placed far apart \cite{kantor1,kantor2}. Scaling theories have provided a deeper insight into the structure of pearl necklaces \cite{avd, avd1, helmut1, helmut2}. However, there are very few analytical theories, mainly based on the Feynman variational method, that confirm the existence of pearl-necklaces in the absence of salt and counter-ion condensation \cite{thomas3,olvera}. One reason for the limited literature is that analytical models require explicit contributions from chain entropy, short range two-body attractive, three-body repulsive and long-range screened Coulombic interactions. The inclusion of  all these terms is a formidable task in a  theoretical description. However, there are methods like the uniform expansion method, which provide systematic and useful way of looking at the systems with many different length scales \cite{edwards1,edwards2, paul}. 


In an earlier shorter communication, one of us used the uniform expansion method in the absence of salt to establish that the globule-to-rod transition is brought about via intermediate pearl-necklaces, which are metastable \cite{arti1}. The transition was found to be first order in nature. The critical fraction of charges where the first-order transition occured was estimated using Maxwell equal area construction. Although the discontinuous nature of transition was established, the mean size of pearl-necklaces and several  other important length scales in polyelectrolytes could not be calculated. Also, how the presence of salt can change the nature of transition and dissolve pearl-necklace structures was not considered. 

In this work, we present a detailed calculation for polyelectrolytes in poor solvents in the presence and absence of salt.  We start with a model Hamiltonian for polyelectrolytes in poor solvents incorporating entropic, screened electrostatic, two body attractive and three body repulsive terms. We first obtain the mean size of the electrostatic blobs, the thermal blobs, the beads, the strings and pearl necklaces using the uniform expansion method in the absence of salt. The present study exploits the perturbative nature of the uniform expansion method to  estimate the important length scales of a polyelectrolyte in a poor solvent. The Hamiltonian based approach adopted here allows for a complete investigation of the phase transition between different conformational states. The details are presented in Section III.  In the presence of salt, the screened electrostatic interactions are characterized by the Debye screening length. Depending on the ratio of the Debye screening length and the size of the polymer, the nature of rod-to-globular transition changes from discontinuous to continuous.   In Section IV, we study in detail the nature of transition as a function of the salt concentration. We determine the salt concentration which dissolves the pearl-necklace structures and brings about a direct transition.

\section{Model}
The Hamiltonian of an infinitely dilute solution of polyelectrolytes in a poor solvent in the presence of salt is given by
\begin{equation}
{\cal H} = {\cal H}_0 + {\cal H}_{2} + {\cal H}_3 + {\cal H}_c, 
\end{equation}
where
\begin{eqnarray}
{\cal H} &=& \frac{3}{2b^2} \int _{0}^{N} ds \left(\frac{\partial{\bf r}(s)}{\partial s}\right)^2 -
\frac{v}{2} \int_{0}^{N} ds \int_{0}^{N} ds^{\prime} \delta[{\bf r}(s) - {\bf r}(s^\prime)] 
+ \frac{w}{6} \int_{0}^{N} ds \int_{0}^{N} ds^{\prime} \int_{0}^{N} ds^{\prime\prime}\delta[{\bf r}(s) - {\bf r}(s^\prime)] \delta[{\bf r}(s^\prime) - {\bf r}(s^{\prime{\prime}})]\nonumber\\ 
& & + \frac{v_{c}}{2} \int_{0}^{N} ds \int_{0}^{N} ds^{\prime}  \frac{e^{-\kappa|{\bf r}(s) - {\bf r}(s^{\prime})|}}{|{\bf r}(s) - {\bf r}(s^{\prime})|} , 
\end{eqnarray}
where the polymer conformation is decribed by the radius vector ${\bf r}(s)$ at distance $s$ from one end of the chain; $b$ is the Kuhn length and $N$ is the number of monomers. ${\cal H}_0$ describes the entropic elasticity of the chain; ${\cal H}_2, {\cal H}_3$ and ${\cal H}_c $ account for the two-body attractive,  three-body repulsive and the screened coloumbic interactions respectively;  $v$, $w$ and $v_c$ represent the strengths of the two-body, three-body and electrostatic interactions respectively. The temperature dependence of $v$ is given by $v = \tau b^3$, where $\tau = |T - \theta|/\theta$ is the reduced temperature and $\theta$ is the theta temperature \cite{arti1}. The sign of the two-body interaction term is positive under good solvent conditions and negative under poor solvent conditions. $\kappa$ is the inverse screening length which depends on the ionic strength by $r_D^{-2} = \kappa^2 = 4\pi l_B \sum_i c_i q_i^2$, where $r_D$ is the Debye screening length; $c_i$ is the concentration of the $i$th ion with charge $q_i$; $l_B = e^2/\epsilon k_B T$ is the Bjerrum length defined as the distance at which the electrostatic interaction between the two elementary charges is equal to the thermal energy $k_B T$ and $\epsilon$ is the dielectric constant of the medium. The strength of the electrostatic interaction is given by $v_c = f^2 l_B$, where $f$ is the fraction of charges along the chain backbone. It is to be noted that the fraction of charges are assumed to be too small to bring about counterion condensation.

In terms of the Fourier transform, the interaction energy can be written as
\begin{eqnarray}
{\cal H}_2 &=& -\frac{v}{2} \int_{0}^{N} ds \int_{0}^{N} ds^{\prime} \int\frac{d^3{\bf k}}{(2\pi)^3}~e^{i{\bf k}\cdot|{\bf r}(s) - {\bf r}(s^\prime)|}\\
{\cal H}_c &=& 4\pi \frac{v_{c}}{2} \int_{0}^{N} ds \int_{0}^{N} ds^{\prime}  \int\frac{d^3{\bf k}}{(2\pi)^3}~\frac{e^{i{\bf k}\cdot|{\bf r}(s) - {\bf r}(s^\prime)|}}{k^2+\kappa^2}\\
{\cal H}_3 &=&  \frac{w}{6} \int_{0}^{N} ds \int_{0}^{N} ds^{\prime} \int_{0}^{N} ds^{\prime\prime} \int\frac{d^3{\bf k}}{(2\pi)^3} \int\frac{d^3{\bf q}}{(2\pi)^3}~e^{i{\bf k}\cdot|{\bf r}(s) - {\bf r}(s^\prime)|}~e^{i{\bf q}\cdot|{\bf r}(s^\prime) - {\bf r}(s^{\prime\prime})|}
\end{eqnarray}
The calculation of the mean size of the polyelectrolyte using the above Hamiltonian is a difficult task. A self-consistent perturbative approach of Edwards and Singh known as the uniform expansion method can be used to calculate the polymer size \cite{edwards1,edwards2,paul}. The uniform expansion method is based on defining a new step length, $b_1 \gg b$, such that the mean square end-to-end distance of the chain in presence of the excluded volume remains Gaussian and is governed by $\left< {\bf R}^2 \right> = N b_1^2$. This can only happen if the original Hamiltonian ${\cal H}_{0} = \frac{3}{2b^2} \int _{0}^{N} ds {\dot{\bf r}}(s)^2$ is replaced with the reference Hamiltonian ${\cal H}_{1} = \frac{3}{2b_1^2} \int _{0}^{N} ds {\dot {\bf r}}(s)^2$.
This step can be carried out by adding and subtracting ${\cal H}_{1}$ from ${\cal H}_{0}$ to give the following expression for the mean square end-to-end distance:
\begin{equation}
\label{Ham}
\left< {\bf R}^2 \right> = 
\frac{\int {\cal D}[r(s)]{\bf R}^2 \exp[-({\cal H}_{1}+ 
({\cal H}_{0}-{\cal H}_{1})+{\cal H}_{2}+{\cal H}_3 + {\cal H}_c)]}{\int {\cal D}[r(s)] 
\exp[-({\cal H}_{1}+ ({\cal H}_{0}-{\cal H}_{1})+{\cal H}_{2}+{\cal H}_3 + {\cal H}_c)]},
\end{equation}
One can now expand
$\left< {\bf R}^2 \right>$ in a perturbative series about the reference
Hamiltonian ${\cal H}_{1}$ such that to the first order correction in $v$ one obtains the following variational equation for the unknown parameter $b_1$:
\begin{equation}
\label{eqn1}
\left< {\bf R}^2 \right>_{1} \left< {\cal H}_{0}-{\cal H}_{1}+{\cal H}_2+ {\cal H}_3 + {\cal H}_c \right>_{1} - \left< {\bf R}^2 ({\cal H}_{0}-{\cal H}_{1}+{\cal H}_2+ {\cal H}_3 + {\cal H}_c) \right>_{1} = 0,
\end{equation}
where the average is with respect to the following 
probability distribution: 
$$
{\cal P}[{\bf r}(s)] \propto \exp \left[{-\frac{3}{2b_1^2} \int _{0}^{N} ds
    {\dot{\bf r}}(s)^2}\right].
$$  
Since the probability
distribution is Gaussian, the evaluation of the averages in Eq. (\ref{eqn1}) is simple. The
details of the calculation are presented in Appendix A; here we
simply present the result:
\begin{equation}
\label{eqn3}
1 - \alpha^2 - \frac{ \tau N^{1/2}}{\alpha^3} + \frac{f^2 l_B N^{3/2} F(\beta)}{b\alpha\beta^{5/2}} + \frac{1}{\alpha^6} = 0.
\end{equation}
In the above equation, $\alpha = R/R_{0}$ is the expansion (or compression) factor, which characterizes the change in the polymer size with respect to the unpertubed size $R_{0} = N^{1/2}b$;  $v_c = f^2 l_B $; $v = \tau b^3$; $w=b^6$; $\beta = \kappa^2 Nb_1^2/6 = \left<{R}_g^2 \right>/r_D^2$, where $\beta$ represents the dimensionless ratio of the mean square polymer size with respect to the square of the Debye screening length $r_D$. $F({\beta})$ is given by    
\begin{equation}
\label{eqn4}
F(\beta) = e^\beta [1-\phi(\sqrt\beta)] (1-2\beta/3 +\beta^2/6) + 2\sqrt{\beta /\pi }-\beta/3 -1,
\end{equation}
where $\phi$ is the probability integral. The first two terms in Eq. (\ref{eqn3}) account for the entropy of the
polymer; the third term is due to the excluded volume interactions between chain segments which is attractive in nature; the fourth term accounts for the electrostatic repulsion; the last term represents the three-body interactions. For convenience, we have ignored all the numerical coefficients.

\section{Polyelectrolyte chain in a salt-free solution}

The limit of $\beta \rightarrow 0$ in Eq. (\ref{eqn4}) represents the case of a salt-free solution. In this limit $F(\beta)/\beta^{5/2} \rightarrow 1 $ 
and Eq. (\ref{eqn3}) reduces to
\begin{equation}
\label{eqn5}
1 - \alpha^2 - \frac{ \tau N^{1/2}}{\alpha^3} + \frac{f^2 l_B N^{3/2}}{b\alpha} + \frac{1}{\alpha^6} = 0,
\end{equation}
The above equation was solved analytically in Ref. \cite{arti1} in two opposite limits. In a poor solvent at low fraction of charges, $\alpha \ll 1$, $\tau N^{1/2} \gg 1$  and $f^2 l_{B}N^{3/2}/b \ll 1$, the attractive interactions dominate, and the comparison of the third and the last term gives a globule of size $R_{glob} \sim b \tau^{-1/3} N^{1/3}$.  At large fraction of charges,  $\alpha \gg 1$, $\tau N^{1/2}\ll 1$ and $f^2l_{B} N^{3/2}/b \gg 1$, the electrostatic repulsion overcomes the attractive interaction. As a result, the comparison of the second and the fourth term  gives $\alpha = (f^2 l_{B}/b)^{1/3} N^{1/2}$ resulting in a rod-like state, $R_{rod} \sim bN(f^2l_{B}/b)^{1/3}$. Although the approach  used in Ref. \cite{arti1} gives the right prediction for the globular and rod-like state, it can not predict the size of pearl necklaces. This is because the formation of pearl necklaces involves contributions from entropic, steric and electrostatic interactions, and therefore requires comparison of more than two terms.

Strictly speaking, the uniform expansion method is a perturbative approach, where all the interaction terms other than the entropic contribution should be considered as a perturbation \cite{edwards1,edwards2,paul}. Such an approach can compare more than two terms analytically. In what follows, we make use of the perturbative nature of the uniform expansion method to calculate the local length scales \cite{paul}, which are then used to estimate the pearl necklace size. In terms of the blob picture, the above scaling can be reconsidered in the following limiting cases. 

\subsection{Polyelectrolytes in poor solvents at low fraction of charges}

At low fraction of charges, the electrostatic interaction term (the fourth term in Eq.(\ref{eqn5})) can be ignored. In this limit, an important local length scale is the thermal blob size \cite{degennes}. For sizes less than the thermal blobs $\xi_T$, monomers do not feel the attractive interactions due to the poor solvent conditions and follow gaussian statistics, i.e., $\xi_T \sim N_T^{1/2} b$, where $N_T$ is the number of monomers in the thermal blob. For sizes larger than $\xi_T$, the thermal blobs are collapsed to form a compact sphere, i.e., $R_{glob} \sim (N/N_T)^{1/3} \xi_T$. To get the number of monomers in the thermal blob, let us first consider  a region close to the $\Theta$ temperature, where $\tau \sim 0$. In this region, Eq. (\ref{eqn5}) suggests that $\alpha \sim 1$, and the chain follows Gaussian statistics with size given by $R_0 \sim N^{1/2} b$.  As the solvent quality becomes poor, the effects of the attractive interactions are realized at long length scales. However, at short length scales the chain remains ideal with the size dictated by $\xi_T$. This implies that the excluded volume term (the third term in Eq. (\ref{eqn5})) is a perturbation to the gaussian size $\alpha \sim 1$ as long as $\tau N^{1/2} \ll 1$. The latter gives the number of monomers $N_T \sim 1/\tau^2$ inside the thermal blob of size given by
\begin{equation}
\xi_T \sim N_T^{1/2} b \sim b/\tau
\end{equation}
For $N<N_T$, the statistics is gaussian and the size is dictated by $\xi_T$. For $N > N_T$, the attractive nature of the interaction term dominate and the thermal blobs are space filling with the size given by $R_{glob} \sim (n_T)^{1/3} \xi_T$, where $n_T = N/N_T$ are the number of thermal blobs.  The overall size scales as  
\begin{equation}
R_{glob} \sim b \tau^{-1/3} N^{1/3},
\end{equation}
which recovers the size of a polymer globule at  low fraction of charges.

\subsection{Polyelectrolytes in poor solvent at large fraction of charges}

At large fraction of charges, the electrostatic interactions dominate over the attractive interactions, and the latter can be ignored. This limit corresponds to theta solvent, where electrostatic blobs are linearly arranged. In this limit, an important local length scale is the electrostatic blob size, $\xi_{elec}$. For sizes less than the electrostatic blobs, $\xi_{elec}$, monomers do not feel the repulsive  electrostatic interactions and follow gaussian statistics, i.e., $\xi_{elec} \sim N_{elec}^{1/2} b$, where $N_{elec}$ is the number of monomers in the electrostatic blobs. For sizes larger than $\xi_{elec}$, the electrostatic interactions dominate and blobs are linearly arranged to result in an extended state, i.e., $R_{rod} \sim (N/N_{elec}) \xi_{elec}$. In this limit, the electrostatic interaction term (the fourth term in Eq. (\ref{eqn5})) is a perturbation to the gaussian size $\alpha \sim 1$ as long as $f^2 l_B N^{3/2}/b \ll 1$. Thus the number of monomers $N_{elec}$ that retain the gaussian statistics are given by
\begin{equation}
N_{elec} \sim (f^2 l_B/b)^{-2/3}, 
\end{equation}
such that for $N < N_{elec}$ the statistics is ideal with the size $\xi_{elec} \sim N_{elec}^{1/2} b$ given by 
\begin{equation}
\xi_{elec} \sim (f^2 l_B/b)^{-1/3} b.
\end{equation}
Therefore, a short chain follows gaussian statistics. For $N > N_{elec}$, the electrostatic blobs are arranged linearly to give the mean size $R_{rod} = (N/N_{elec}) \xi_{elec}$, where $n_{elec} = N/N_{elec}$  are the number of electrostatic blob at a given fraction of charges. Therefore, the overall size is given by
\begin{equation}
R_{rod} \sim bN(f^2l_{B}/b)^{1/3}, 
\end{equation}
which recovers the polyelectrolyte rod size at high fraction of charges. In the limit where $f \sim (l_B/b)^{-1/2} N^{-3/4}$, the polyelectrolyte rod size $R_{rod}$ crosses over to the gaussian size $R_0$.

\subsection{Polyelectrolytes in poor solvents at intermediate fraction of charges}
 
To understand the behavior in the intermediate regime, Eq. (\ref{eqn5}) can be solved numerically. The numerical solution is shown in Fig. (1), which is a plot of the dimensionless size $\alpha$ as a function of the dimensionless electrostatic energy $\overline{f} = f^2 l_B N/b$. The figure shows an intermediate regime where the curve folds back on itself. This region represents a bistable equilibrium between the collapsed globules and elongated strings that can coexist. This is the region where pearl-necklace structures are formed. In the intermediate region between $\overline{f}_{min}$ and $\overline{f}_{max}$, there are three curves: the two curves increases with the increase in the fraction of charges and represent the metastable pearls and necklaces; one curve decreases with the increase in the fraction of charges and represent the unstable region.  

 A detailed analysis of this plot was presented in Ref. \cite{arti1}, where the Maxwell equal-area construction was used to determine the critical fraction of charges where the abrupt transition  (spontaneous Rayleigh splitting) occured. The pearl-necklace structures formed at the intermediate fraction of charges are the result of the balance between the long range repulsive electrostatic interactions and the short range attractive interactions. In terms of the above scaling, the intermediate regime corresponds to $\overline{f} = f^2 l_B N/b \sim \tau$.  This gives the critical fraction of charges, $f_c \sim (\tau b/ l_B N)^{1/2}$, where the abrupt transition is expected to take place. For $\tau = 1.0$ in Fig. (1), the scaling predicts the transition at $\overline{f}_c = f_c^2 l_B N/b \sim 1.0$. This is very close to the value of $\overline{f}_c = 0.9$ obtained  from the Maxwell equal area construction used in Fig. (1). Therefore, at the intermediate fraction of charges,  the number of monomers inside a Rayleigh blob \cite{burkhard} can be obtained by comparing the third and the fourth terms in Eq. (\ref{eqn5}). Since the statistics at the local length scale is gaussian, $\alpha \sim 1$, the number of monomers inside the blob are given by
\begin{equation}
\label{ray}
N_{R} \sim \tau b/f^2 l_B. 
\end{equation}
A globule splits into smaller globules as soon as the electrostatic energy, $k_B T f^2 l_B N^2/ n_T^{1/3} \xi_T$, overcomes the surface energy, $k_B T n_T^{2/3}$ \cite{burkhard}. The balance of the two terms gives the number of monomers in a Rayleigh blob, $N_R \sim \tau b/ f^2 l_B$. The latter expression is the same as Eq. (\ref{ray}). This is because the surface energy considerations are already there implicitly in the present formalism in terms of the fundamental interactions.    

It is clear from Fig. (1) that in the intermediate regime between $\overline{f}_{min}$ and $\overline{f}_{max}$, the size of a pearl-necklace structure  is determined by the size of the pearls (beads $d_{bead}$) and necklaces (strings $l_{str}$) that contribute to the overall structure. The size of the pearl-necklace structure can be calculated in two limits: when $d_{bead} \gg l_{str}$, the pearl-necklace structure is dominated by beads and corresponds to the regime $\overline{f}_{min} \simeq \overline{f} < \overline{f}_c$, represented by a dashed line on the left hand side of $\overline{f}_c$; the opposite limit of  $l_{str} \gg d_{bead}$ corresponds to the regime $\overline{f}_c < \overline{f} \simeq \overline{f}_{max}$. It is worth mentioning that the size of pearl-necklaces follow universal scaling in terms of $N/N_{R}$ only close to the globular and rod-like states. This is because the globular and rod-like regions are separated by a first-oder transition \cite{burkhard}. 

In what follows, the size of pearl-necklaces is estimated in the limits mentioned above. With the increase in the fraction of charges, the collapsed globule with size given by $R_{glob} \sim (N/N_T)^{1/3} \xi_T$ is expected to break down into small sized globules (beads), the size of which is given by  
\begin{equation} 
d_{bead} \sim (N_{R}/N_T)^{1/3} \xi_T \sim N_{R}^{1/3} \tau^{-1/3} b \sim (b/f^2 l_B)^{1/3} b.  
\end{equation}
It is clear from Fig. (1) that for $f_{min} \simeq f < f_c$, the size of the bead $d_{bead}$ is much larger than the length of the string $l_{str}$, {\it{i.e.}}, $d_{bead} \gg l_{str}$.  In this limit, for $N > N_{R}$ the electrostatic interactions determine the overall size 
\begin{equation}
\label{pn1}
R_{p} \sim (N/N_{R}) d_{bead} \sim (f^2l_{B}/b)^{2/3} bN/\tau. 
\end{equation}
This gives the size of a pearl-necklace chain in the limit $d_{bead} \gg l_{str}$. The size obtained in this limit is similar to the elongated globule picture of Khokhlov \cite{khokh}. It is to be noted that the number of pearls are given by $n_p \sim N/N_R\sim (f^2 l_B/\tau b) N$. By changing the fraction of charges $f$ or $\tau$, a globule splits into different (integer) number of pearls, $n_p$, resulting in a cascade of abrupt transitions between necklace globules. In References \cite{burkhard, avd, avd1}, a phase diagram between the fraction of charges $f^2 l_B/b$ and $\tau$ show a cascade of transitions with different number of pearls, which occur because of the change in either $f$ or $\tau$ . When a globule splits up into progressively large number of smaller globules (pearls), the total size of the pearls decreases while the total length of the strings increases.  In the context of present work, the region between $\overline{f}_{min}$ and $\overline{f}_{max}$ represents a cascade of transitions, where  a globule splits into different number of pearls sizes as soon as $\overline{f} > \overline{f}_{min}$. The decreases in the overall size of the pearls is a reflection of the increase in its number. The intermediate regions in Figs. (1) and (3a) represent two such instances where a cascade of transitions occur because of the change in $f$ and $\tau$ respectively.

In the limit $\overline{f}_{c} < \overline{f} \simeq \overline{f}_{max}$, the length of the string $l_{str}$ is much larger than the size of the string $d_{bead}$, {\it{i.e.}}, $l_{str} \gg d_{bead}$. On the local length scale, the length of the string is dictated by the gaussian statistics given by 
\begin{equation} 
\label{neck}
l_{str} \sim (N_{R}/N_T)^{1/2} \xi_T \sim (\tau b/f^2 l_B)^{1/2} b.
\end{equation}
The above scaling can also be obtained be comparing the surface energy of the string, $k_B T l_{str}/ \xi_T$, with the electrostatic repulsion between the pearls, $k_B T f^2 l_B N_R^2 / l_{str}$ \cite{burkhard}. On large length scale, the electrostatic interactions dominate and the overall size is given by
\begin{equation}
\label{pn2}
R_{n} \sim (N/N_{R}) l_{str} \sim (f^2 l_B/b \tau)^{1/2} bN, 
\end{equation}
which yields the pearl-necklace size in the limit $l_{str} \gg d_{bead}$. 

The size of the bead, string, pearl-necklace and the number of pearls calculated from the uniform expansion method is in complete agreement with the previous scaling theories \cite{avd,avd1, helmut1,helmut2}. The advantage of the present theory is that one can easily  include the effects of salt. The details are presented in the next section.

\section{Polyelectrolyte chain in the presence of salt}
The limit of $\beta \gg 1$ in Eq. (\ref{eqn4}) represents the case of high salt concentration. In this limit $F(\beta) \rightarrow \beta^{3/2} $ 
and Eq. (\ref{eqn3}) reduces to
\begin{equation}
\label{eqn6}
1 - \alpha^2 - \frac{ \tau N^{1/2}}{\alpha^3} + \frac{f^2 l_B N^{3/2}}{b\alpha\beta} + \frac{1}{\alpha^6} = 0,
\end{equation}
Since $\beta = \kappa^2 N b_1^2 = \kappa^2 \alpha^2 N b^2$, the above equation reduces to
\begin{equation}
\label{eqn7}
1 - \alpha^2 - \frac{ \tau N^{1/2}}{\alpha^3} + \frac{f^2 l_B N^{1/2}}{b^3\alpha^3 \kappa^2} + \frac{1}{\alpha^6} = 0,
\end{equation}
In a theta solvent, the electrostatic term is a perturbation ($\alpha \sim 1$) as long as ${f^2 l_B N^{1/2}}/{b^3 \kappa^2}  \ll 1$. This gives the $\kappa$ dependent electrostaic blob size $\xi_{el,salt} \sim N_{el,salt}^{1/2} b$, where
\begin{equation}
N_{el,salt} \sim (b^3 \kappa^2/f^2 l_B)^2,
\end{equation}
is the number of monomers inside the electrostatic blob of size given by
\begin{equation}
\xi_{el,salt} \sim b^4 \kappa^2/f^2 l_B.
\end{equation}
In a theta solvent, the third term in Eq. (\ref{eqn6}) can be ignored and the fourth term represent the excluded-volume type interaction term. In this limit, the comparison of the second and the fourth terms yields $R_{sa, salt} \simeq (f^2 l_B/b^3 \kappa^2)^{1/5} N^{3/5} b$. The latter represents a self avoiding walk with an effective persistence length given by 
\begin{equation}
l_{p,salt} \sim  (f^2 l_B/b^2 \kappa^2)^{1/5} b. 
\end{equation}
With the increase in the salt concentration the $\kappa$-dependent electrostatic  blob size will increase until $\xi_{el,salt}$ becomes of the order of  $\xi_{elec}$.  The comparison of the two gives the minimum value of the salt concentration that influences the electrostatic blob size,
\begin{equation}
\kappa_{min} \sim (f^2 l_B/b)^{1/3} b^{-1}. 
\end{equation}
 For $\kappa < \kappa_{min}$, the charges on the chain backbone are unscreened and the size is given by $R_{rod} \sim bN(f^2l_{B}/b)^{1/3}$. In the opposite limit, $\kappa > \kappa_{min}$, some of the charges are screened and the size is given by $R_{sa, salt} \simeq (f^2 l_B/b^3 \kappa^2)^{1/5} N^{3/5} b$. 

The blob size increases with increase in salt concentration until $\kappa = \kappa_{\theta}$. At this salt concentration, the effective excluded volume, $ v^* = \tau - f^2 l_B/b^3 \kappa^2$, becomes zero and 
\begin{equation}
\label{salt}
\kappa_{\theta} \sim (f^2 l_B/b^3 \tau)^{1/2}. 
\end{equation}
At $\kappa = \kappa_{\theta}$, the electrostatic blob size $\xi_{el,salt}$ becomes equal to the thermal blob size $\xi_T = b/\tau$ and the chain behaves like an ideal chain. For $\kappa > \kappa_{\theta}$, the effective excluded volume becomes negative due to which the attractive interaction dominates and  the thermal blobs of size $\xi_T$ collapse to form a globule of size $R_{glob} \sim b \tau^{-1/3} N^{1/3}$. Thus the pearl-necklace intermediates are dissolved at the salt concentration given by $\kappa = \kappa_{\theta}$. 

The comparison of Eq. (\ref{salt}) with Eq.(\ref{neck}) shows that $\kappa_{\theta} \sim 1/l_{str}$ implying $r_D \sim l_{str}$. The latter shows that the salt concentration that dissolves the pearl necklace structure corresponds to the case where the Debye screening length is of the order of the necklace (string) size.

To understand these results quantitatively, Eqs. (\ref{eqn3}) and (\ref{eqn4}) can be solved numerically. The result is presented in Figure 2, which is a plot of the dimensionless ratio of the chain size $\alpha$ as a function of the dimensionless inverse sreening length $\beta$. The curve shows a slow decrease in the chain size with the increase in the salt concentration followed by an abrupt collapse to a globular state. As suggested by the above scaling arguments, the abrupt transition is expected to take place at  salt concentration given by $\kappa_{\theta} \sim (f^2 l_B/b^3 \tau)^{1/2}$. In terms of the dimensionless parameters $\beta$, the transition is expected to take place when $\beta_c = \kappa_\theta^2 N b^2 /6$. For $\tau = 0.5$, $f^2 l_B/b = 0.1$ and $N = 100$, the scaling predicts the transition at $\beta_c = 3.3$. This is very close to the value of  $\beta_c = 3.0$ in Fig. (2) obtained by solving Eqs. (\ref{eqn3}) and (\ref{eqn4}) numerically.

Figures 3a-3c are the plots of the dimensionless size $\alpha$ as a function of  the reduced temperture $\tau$  at three different salt concentrations characterized by $\kappa b = 0.17, 0.35, 0.42$ respectively, obtained by solving Eqs. (\ref{eqn3}) and (\ref{eqn4}) numerically. For the values of $N = 100$, $f^2l_B/b = 0.05$ used in Fig. 3a-3c, the minimum salt concentration required to influence the polyelectrolyte size is given by $\kappa_{min}  b \sim (f^2 l_B /b )^{1/3} \sim 0.37$. Fig. 3a corresponds to the case where $\kappa < \kappa_{min}$. As a result, the charges on the polyelectrolytes are completely unscreened and the first-order transition proceeds via the formation of an intermediate pearl-necklace structure; Figure 3b represents the intermediate regime where $\kappa \sim \kappa_{min}$ and some of the charges on the chain are screened. In this limit the fraction of charges on the chain are too small to form the metastable pearl-necklace structure and the transition is weakly first order. As before, the transition is expected to take place when $\kappa \sim \kappa_\theta \sim (f^2 l_B /\tau b^3)^{1/2}$. For $f^2 l_B/b = 0.05$ and $\kappa b = 0.35$, the scaling suggests that the  transition takes place at $\tau_c \sim 0.4$. This is very close to the value of $\tau_c = 0.37$ in Fig. 3b; Figure 3c represents the case where $\kappa > \kappa_{min}$ and the charges on the chain backbone are completely screened. In this limit the transition becomes smooth and continuous second-order in nature. 

\section{Conclusions}
      
We have used the uniform expansion method of Edwards and Singh to study conformational transitions of polyelectrolytes in  poor solvents in the presence and absence of salt.  The presence of the additional long-range repulsion due to the electrostatic interactions between charged monomers destabilizes the globular structure as the fraction of charged monomers along the chain backbone is increased and leads to the formation of pearl-necklaces at intermediate fraction of charges . The uniform expansion method allows us to determine the the size of the pearls (beads), necklaces (strings) and pearl-necklace intermediate, and shows a first order transition between a globular state to an extended state via metastable pearl-necklaces.  


In the presence of salt, the electrostatic blob size depends on the salt concentration and is found to be larger than that of the unscreened case. We determine the minimum salt concentration, $\kappa_{min}$, at which the size of the electrostatic blob is influenced. For $\kappa < \kappa_{min}$, the size of the polyelectrolyte rod is governed by the unscreened electrostatic interactions.  For $\kappa \sim \kappa_{min}$, on the other hand, some of the charges are screened and this results in the increase in the electrostatic blob size amounting to decrease in the polyelectrolyte size. The electrostatic blob size increases with the increase in the salt concentration until it becomes equal to the thermal blob size.  We determine the salt concentration, $\kappa_{\theta}$, at which the thermal blob size is equal to the electrostatic blob  and the polyelectrolyte behaves like a neutral ideal polymer.  This is the concentration at which the pearl-necklace intermediates are dissolved and the transition proceeds directly. For $\kappa > \kappa_{\theta}$, most of the charges on the chain backbone are screened and the attractive interactions dominates resulting in the chain collapse to form a globule. Our scaling predicts that the pearl-necklace intermediates dissolve when the Debye screening length is of the order of $l_{str}$.
In a future publication,  the present formalism will be extended to include the effects of counterion condensation \cite{muthu2}.

\appendix
\section{Uniform expansion method}

The detailed calculations of the terms containing the entropic term,  $ \left< {\bf R}^2 \right>_{1} \left< {\cal H}_{0} -{\cal H}_{1} \right>_{1} - \left< {\bf R}^2 ({\cal H}_{0} - {\cal H}_{1})\right>_{1}$ and the two-body interaction, $\left< {\bf R}^2 \right>_{1} \left< {\cal H}_2 + {\cal H}_c) \right>_{1} - \left< {\bf R}^2 ({\cal H}_2 + {\cal H}_c) \right>_{1} $, are fairly standard and can be found in References \cite{edwards1} and \cite{edwards2}. 
\begin{eqnarray}
\left< {\bf R}^2 \right>_{1} \left< {\cal H}_{0} -{\cal H}_{1} \right>_{1} - \left< {\bf R}^2 ({\cal H}_{0} - {\cal H}_{1})\right>_{1} &=& N{b_{1}} ^4 \left(\frac{1}{{b_1}^2} - \frac{1}{b^2}\right),\\
\left< {\bf R}^2 \right>_{1} \left< {\cal H}_{2} \right>_{1} - \left< {\bf R}^2 ({\cal H}_{2})\right>_{1} &=& - \frac{2 v {b_1}^4}{9 (2\pi)^2} \int_{0}^{\infty}dk\int_{0}^{N} ds \int_{0}^{s} ds^{\prime} (s-s^{\prime})^2 k^4 e^{-k^2{b_1}^2(s-s^{\prime})/6},\\
\left< {\bf R}^2 \right>_{1} \left< {\cal H}_{c} \right>_{1} - \left< {\bf R}^2 ({\cal H}_{c})\right>_{1} &=& \frac{{8\pi v_c} {b_1}^4}{9 (2\pi)^2 b^2} \int_{0}^{\infty}dk\int_{0}^{N} ds \int_{0}^{s} ds^{\prime}(s-s^{\prime})^2 \frac{k^4}{(k^2 +\kappa^{2})} e^{-k^2{b_1}^2(s-s^{\prime})/6},
\end{eqnarray}

The uniform expansion method of Edwards and Singh does not include the three body interaction term. A detailed calculation of the three body interaction term is given in Ref. 23.  In what follows, we present a few important steps required to calculate the three body interaction term, $\left< {\bf R}^2 \right>_{1} \left< {\cal H}_{3} \right>_{1} - \left< {\bf R}^2 ({\cal H}_{3})\right>_{1}$. Let us first calculate $\left< {\bf R}^2 {\cal H}_3 \right> _1 $,
\begin{equation}
\left< {\bf R}^2 {\cal H}_3 \right>_1 = \frac{w}{6} \int_{0}^{N} ds \int_{0}^{N} ds^{\prime} \int_{0}^{N} ds^{\prime\prime} \left< R^2 \delta[{\bf r}(s) - {\bf r}(s^\prime)] \delta[{\bf r}(s^\prime) - {\bf r}(s^{\prime\prime})] \right>
\end{equation}
In terms of the wave vectors ${\bf k}$ and ${\bf q}$, the above equation can be rewritten as
\begin{equation}
\frac{w}{6} \int_{-\infty}^{\infty} \frac{d^3{\bf k} }{(2\pi)^3}\int_{-\infty}^{\infty} \frac{d^3{\bf  q}}{(2\pi)^3}\int_{0}^{N} ds \int_{0}^{N} ds^{\prime} \int_{0}^{N} ds^{\prime\prime} \left< {\bf R}^2 e^{i {\bf k} \cdot |{\bf r}(s) - {\bf r}(s^\prime)|} e^{i {\bf q} \cdot |{\bf r}(s^\prime) - {\bf r}(s^{\prime\prime})|} \right>
\end{equation}

To calculate the quantity of interest, it is important to expand the mean square end-to-end distance and the probability distribution in terms of the internal coordinates. Since the probability distribution is Gaussian, the averages can easily be calculated to give the following expression:
\begin{eqnarray}
\left< {\bf R}^2 {\cal H}_3 \right>_1 &=&  w \int_{-\infty}^{\infty} \frac{d^3{\bf k} }{(2\pi)^3}\int_{-\infty}^{\infty} \frac{d^3{\bf  q}}{(2\pi)^3}\int_{0}^{N} ds \int_{0}^{s} ds^{\prime} \int_{0}^{s^{\prime}} ds^{\prime\prime} \left[N {b_1}^2 -  \frac{(s-s^{\prime})^2 {\bf k}^2 {b_1}^4}{9} - \frac{(s^{\prime}-s^{\prime\prime})^2 {\bf q}^2 {b_1}^4}{9}\right] \nonumber\\
& & ~~~~~~~~~~~~~~~~~~~~~~~~~~~~~~~~~~~~~~~~~e^{-{\bf k}^2  (s-s^{\prime}) {b_1}^2/6} e^{-{\bf q}^2  (s^{\prime}-s^{\prime\prime}) {b_1}^2/6}
\end{eqnarray}
One can repeat the above procedure to calculate $\left< {\bf R}^2 \right>_{1} \left< {\cal H}_{3} \right>_{1} $, which is given by
\begin{equation}
\left< {\bf R}^2 \right>_{1} \left< {\cal H}_{3} \right>_{1} = w \int_{-\infty}^{\infty} \frac{d^3{\bf k} }{(2\pi)^3}\int_{-\infty}^{\infty} \frac{d^3{\bf  q}}{(2\pi)^3}\int_{0}^{N} ds \int_{0}^{s} ds^{\prime} \int_{0}^{s^{\prime}} ds^{\prime\prime} N {b_1}^2 e^{-{\bf k}^2  (s-s^{\prime}) {b_1}^2/6} e^{-{\bf q}^2  (s^{\prime}-s^{\prime\prime}) {b_1}^2/6},
\end{equation}
After substracting Eq. (A7) from Eq. (A6), one obtains
\begin{eqnarray}
\left< {\bf R}^2 \right>_{1} \left< {\cal H}_{3} \right>_{1} - \left< {\bf R}^2 {\cal H}_3 \right>_1 &=&  w \int_{-\infty}^{\infty} \frac{d^3{\bf k} }{(2\pi)^3}\int_{-\infty}^{\infty} \frac{d^3{\bf  q}}{(2\pi)^3}\int_{0}^{N} ds \int_{0}^{s} ds^{\prime} \int_{0}^{s^{\prime}} ds^{\prime\prime} \left[ \frac{(s-s^{\prime})^2 {\bf k}^2 {b_1}^4}{9} + \frac{(s^{\prime}-s^{\prime\prime})^2 {\bf q}^2 {b_1}^4}{9}\right] \nonumber\\
& & ~~~~~~~~~~~~~~~~~~~~~~~~~~~~~~~~~~~~~~~~~e^{-{\bf k}^2  (s-s^{\prime}) {b_1}^2/6} e^{-{\bf q}^2  (s^{\prime}-s^{\prime\prime}) {b_1}^2/6}
\end{eqnarray}
The above equation can be rewritten as
\begin{eqnarray}
\left< {\bf R}^2 \right>_{1} \left< {\cal H}_{3} \right>_{1} - \left< {\bf R}^2 ({\cal H}_{3})\right>_{1} &=& \frac{4 w   {b_1}^4}{9 (2\pi)^4} \int_{0}^{\infty}dk \int_{0}^{\infty} dq \int_{0}^{N} ds \int_{0}^{s} ds^{\prime} \int_{0}^{s^{\prime}} ds^{\prime\prime}  \left[(s-s^{\prime})^2 k^4 q^2 + (s^{\prime} - s^{\prime\prime})^2 k^2 q^4\right]\nonumber \\
& & ~~~~~~~~~~~~~~~~~~~~~~~~~~~~~~~~~~~~~~~~~~~e^{-k^2{b_1}^2(s-s^{\prime})/6} e^{-q^2{b_1}^2(s^{\prime}-s^{\prime\prime})/6}
\end{eqnarray}

The last
integral diverges as $q \rightarrow \infty$.  The divergence can be removed by introducing an
upper cut-off for the wave number $q$, the details of which are discussed in Ref (23). The integrations in Eqs. (A2), (A3) and (A9)  can easily
be carried out, the results when substituted in Eq. (\ref{eqn1}) gives the
following variational equation:
\begin{equation}
\label{eqn2}
N{b_1}^2\left(1-\frac{{b_1}^2}{b^2}\right)  - \frac{\sqrt{6} v  N^{3/2}}{ \pi^{3/2} {b_1}} + \frac{4 \sqrt{6} v_c N^{5/2} b_{1}F(\beta)}{ b^2\beta^{5/2}} +\left({\frac{3}{\pi}}\right)^3\left[\left({\frac{2}{3}}\right)^{3/2} - \frac{\pi^{1/2}}{4}\right]\frac{w N}{{b_{1}}^4} = 0,
\end{equation}
where $F(\beta) = e^\beta [1-\phi(\sqrt\beta)] (1-2\beta/3 +\beta^2/6) + 2\sqrt{\beta /\pi }-\beta/3 -1$; $\phi$ is the probability integral;  $\beta = \kappa^2 N{b_1}^2/6$ and $\kappa$ is the inverse screening length. The above equation can be
written in a dimensionless form by dividing it by $N{b_1}^2$ and defining
$\alpha = b_{1}/b$. The final form of this variational equation is given by
Eq. (\ref{eqn3}).

\eject
\begin{center}
\begin{large}\bf{Figure Captions}\end{large}
\end{center}
\vskip 0.5in

\noindent \textbf{Figure 1.}  Variation of the dimensionless chain size $\alpha$ as a function of the fraction of charges ${\overline f} = f^2 l_{B}N/b$ for $\tau = 1.0$ and $N = 100$. The region between $\overline{f}_{min}$ and $\overline{f}_{max}$ represents a bistable equilibrium between pearls and necklaces. The Maxwell equal area construction has been used to locate $\overline{f}_c$, the critical fraction of charges where spontaneous Rayleigh splitting occurs \cite{arti1}. \\

\noindent \textbf{Figure 2.} Dependence of the dimensionless chain size $\alpha$ on the dimensionless inverse screening length $\beta$. The results are the numerical solution of Eqs. (\ref{eqn3}) and (\ref{eqn4}) for $N = 100, f^2l_B/b = 0.1$ and $\tau = 0.5$.\\
\\
\noindent \textbf{Figure 3.} Variation of the dimensionless chain size $\alpha$ as a function of the reduced temperature $\tau$. The results are the numerical solution of Eqs. (\ref{eqn3}) and (\ref{eqn4}) for $N = 100, f^2l_B/b = 0.05$, and three different values of $\kappa b$ (a) first-order rod-to-globule transition via the intermediate pearl necklace structure at $\kappa b = 0.17$; (b) weak first-order transition without the intermediate structure at $\kappa b = 0.35$; (c) second-order continuous transition at $\kappa b = 0.42$.\\

\eject

\begin{large}\bf{Figure 1}\end{large}

\begin{center}
\begin{figure}
{\includegraphics[width=1.0\hsize]{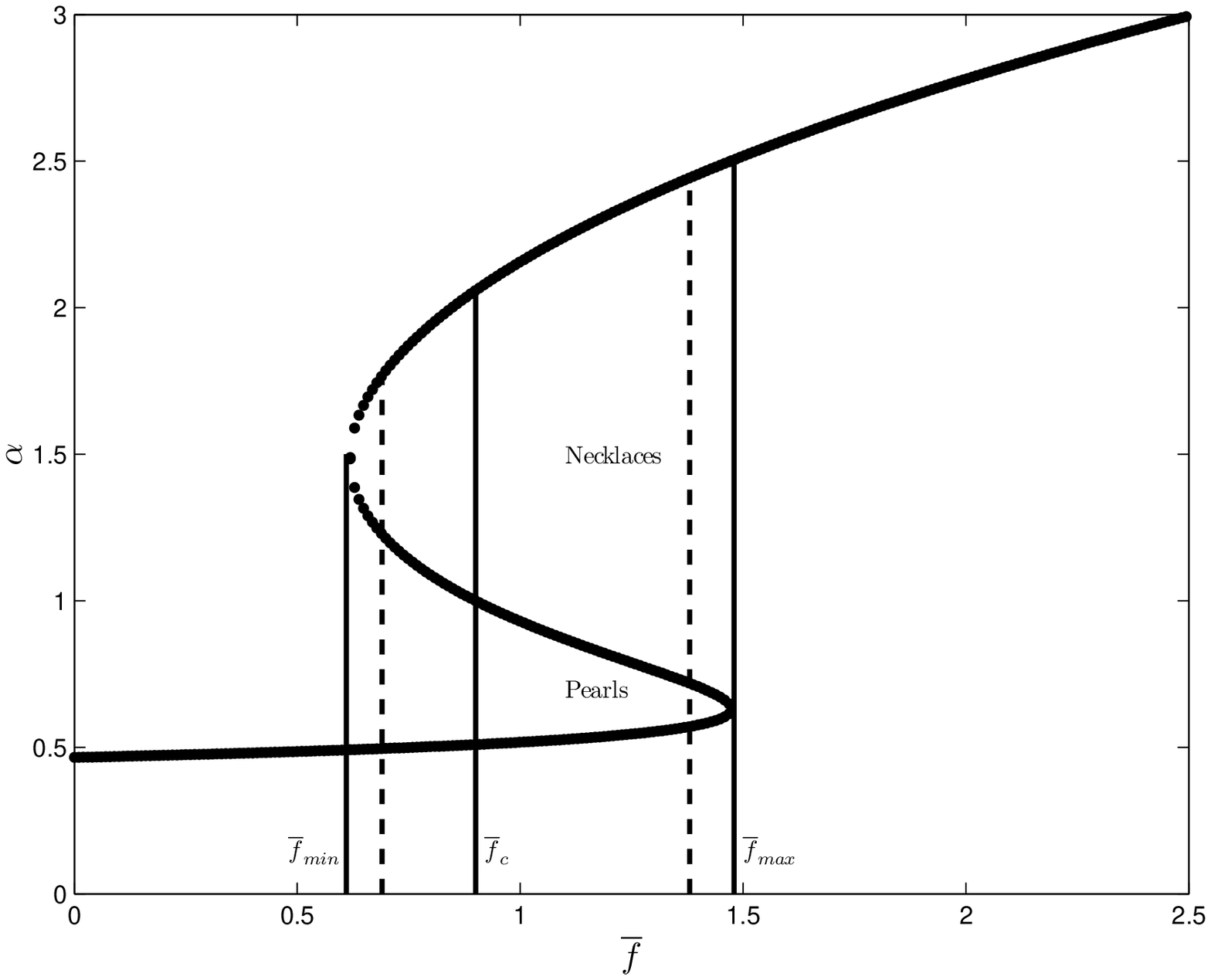}}
\end{figure}
\end{center}
\eject

\begin{large}\bf{Figure 2}\end{large}

\begin{figure}
\begin{center}
\includegraphics[width=1.0\hsize]{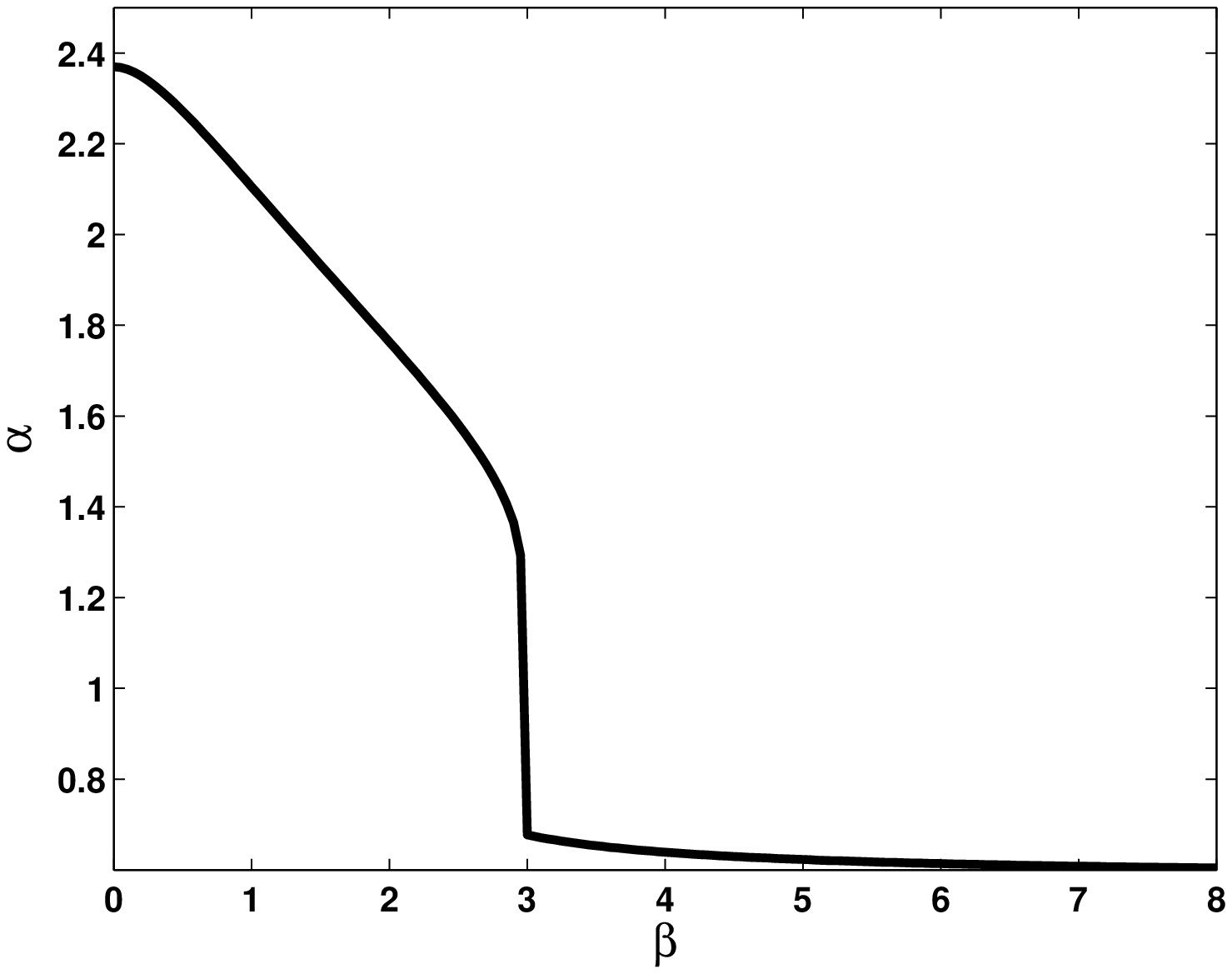} 

\end{center}
\end{figure}

\eject
\begin{large}\bf{Figure 3a}\end{large}
\begin{figure}
\begin{center}
\includegraphics[width=1.0\hsize]{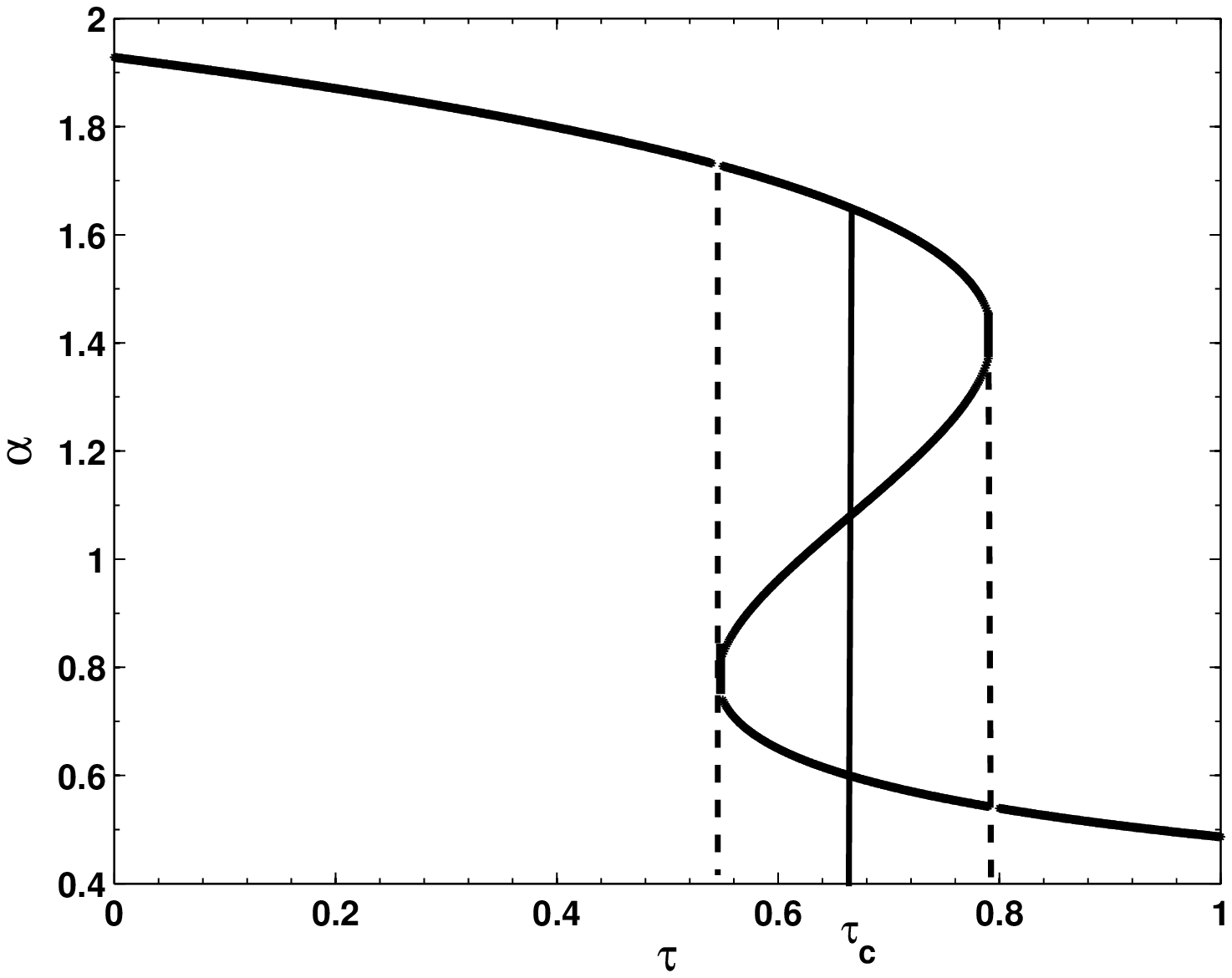} 
\end{center}
\end{figure}

\eject

\begin{large}\bf{Figure 3b}\end{large}
\begin{figure}
\begin{center}
\includegraphics[width=1.0\hsize]{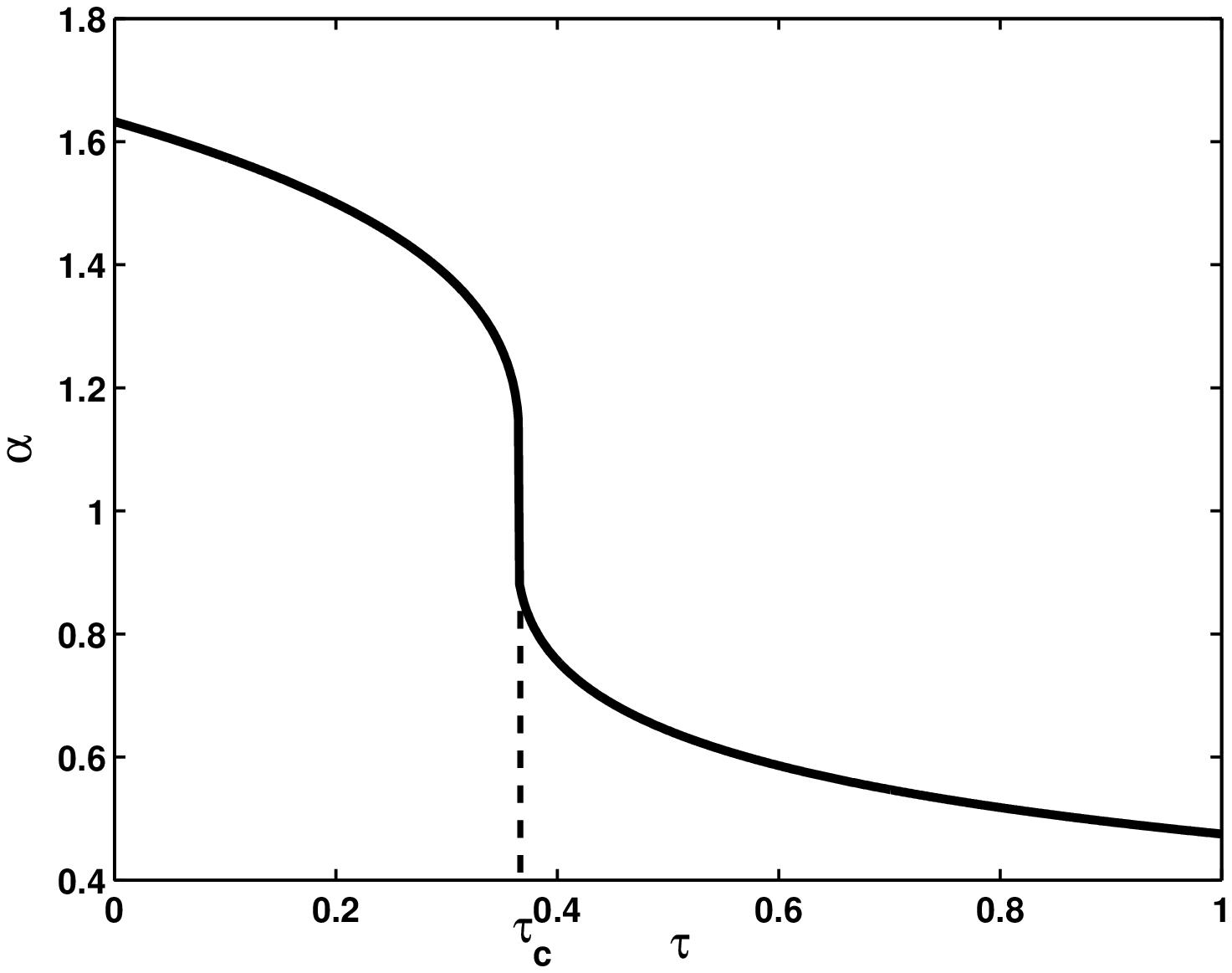} 
\end{center}
\end{figure}

\eject
\begin{large}\bf{Figure 3c}\end{large}

\begin{figure}
\begin{center}
\includegraphics[width=1.0\hsize]{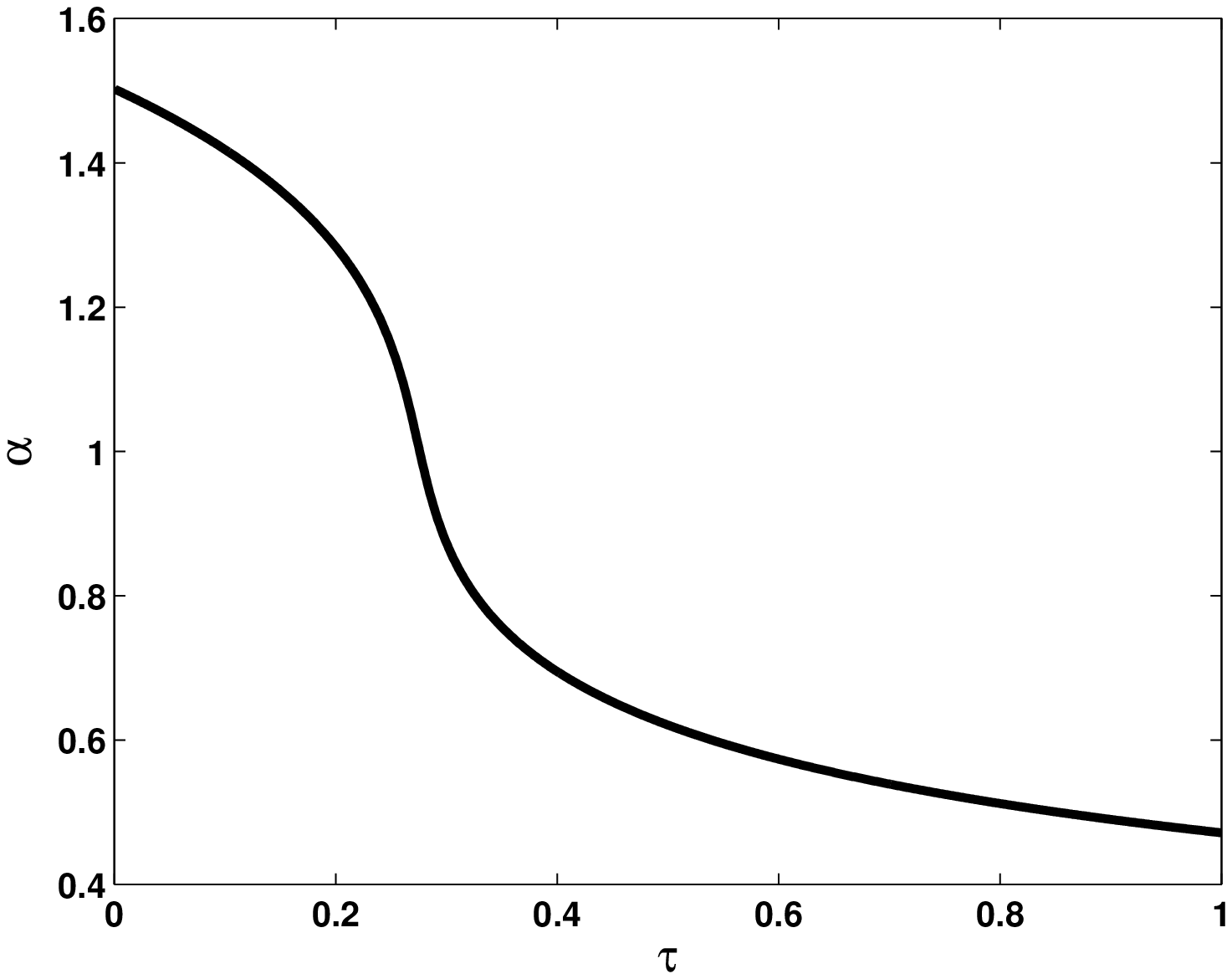} 

\end{center}
\end{figure}


\begin{thebibliography} {99}

\bibitem{t} J.-L Barrat and J-F Joanny,  Adv. Chem. Phys. \textbf{94}, 1 (1996). 

\bibitem{h} M. Hara, Polyelectrolytes; Marcel Dekker: New York, 1993. 

\bibitem{b} K. S. Schmitz, Macroions in solution and colloidal suspension, 1st ed.; VCH Publishers: New York, 1993. 

\bibitem{expt3}  A. Kiriy, G. Gorodyska, S. Minko,  W. Jaeger, P. Stepanek and M. Stamm,  J. Am. Chem. Soc. {\bf 124}, 13454 (2002).

\bibitem{burkhard} A. V. Lyulin,  B. Duenweg, O. V. Borisov and A. A. Darinskii,
Macromolecules \textbf{32}, 3264, (1999). 

\bibitem{stoll} P. Chodanowski and S. Stoll,  J. Chem. Phys. \textbf{111}, 6069  (1999). 

\bibitem{k1} U. Micka and K. Kremer, Europhys. Lett. \textbf{49}, 189 (2000). 

\bibitem{k2} H. J. Limbach and C. Holm,  J. Chem. Phys. \textbf{114}, 9674 (2001). 

\bibitem{k4} C. Holm, H. J. Limbach and K. Kremer,  J. Phys. Condens. Matter \textbf{15}, S205 (2003). 

\bibitem{k3} H. Limbach, C. Holm and K. Kremer,  Europhys. Lett. \textbf{60}, 566 (2002). 


\bibitem{kantor1}  Y. Kantor and M. Kardar, Europhys. Lett. {\bf 27}, 643 (1994).

\bibitem{kantor2} Y. Kantor and M. Kardar, Phys. Rev. E {\bf 51}, 1299 (1995).

\bibitem{avd} A. V. Dobrynin, M. Rubinstein and S. P. Obukhov, Macromolecules {\bf 29}, 2974 (1996).

\bibitem{avd1}  A. V. Dobrynin and M. Rubinstein, Prog. Polym. Sci {\bf 30}, 1049 (2005).

\bibitem{helmut1} H. Schiessel and P. Pincus, Macromolecules \textbf{31}, 7953 (1998). 

\bibitem{helmut2} H. Schiessel, Macromolecules \textbf{32}, 5673 (1999). 



\bibitem{thomas3} G. Migliorini, N. Lee, V. Rostiashvili and T. A. Vilgis, Eur. Phys. J. E  \textbf{6}, 259 (2001). 

\bibitem{olvera} F. J. Solis and  M. Olvera de la Cruz, Macromolecules {\bf 31}, 5502 (1998).




\bibitem{edwards1} M. Doi and  S. F. Edwards, The Theory of Polymer Dynamics; Clarendon Press: Oxford, 1986.

\bibitem{edwards2} S. F. Edwards, P. J. Singh,  Chem. Soc. Faraday Trans. 2, {\bf 75}, 1001 (1979).

\bibitem{paul} P. G. Higgs and J. F. Joanny,  J. Chem. Phys {\bf 94}, 1543 (1990).

\bibitem{arti1} A. Dua and T. A. Vilgis Europhys. Lett. {\bf 71} 49 (2005).

\bibitem{degennes} P. -G. de Gennes,  Scaling Concepts in Polymer Physics; Cornell University Press: Ithaca, NY, 1985.

\bibitem{khokh} A. R. Khokhlov, Journal of Physics A {\bf 13}, 979 (1980).

\bibitem{arti2} A. Dua and T. A. Vilgis Macromolecules {\bf 40}, 6765 (2007).

\bibitem{muthu2} M. Muthukumar, J. Chem. Phys. {\bf 120}, 9343 (2004).






\end{thebibliography}
\end{document}